\documentclass[11pt,a4paper]{article}
\usepackage[utf8]{inputenc}
\usepackage{amsmath,amssymb}
\usepackage{mathtools}
\usepackage{a4wide}
\usepackage{cite}
\usepackage{authblk}
\usepackage{mathrsfs}
\usepackage{color}
\usepackage{url}

\title{Three-generation solutions of equations of motion \\in heterotic supergravity}
\author[1]{Maki Takeuchi\footnote{191s107s@stu.kobe-u.ac.jp}}
\author[2]{Takanao Tsuyuki\footnote{tsuyuki@cc.kogakuin.ac.jp}}
\author[3,4]{Hikaru Uchida\footnote{h-uchida@particle.sci.hokudai.ac.jp}}
\affil[1]{\small \textit{Department of Physics, Kobe University, Kobe 657-8501, Japan}}
\affil[2]{\small \textit{Division of Liberal-Arts, Kogakuin University, Hachioji, Tokyo 192-0015, Japan}}
\affil[3]{\small \textit{Department of Physics, Hokkaido University, Sapporo 060-0810, Japan}}
\affil[4]{\small \textit{Institute for the Advancement of Graduate Education, Hokkaido University, Sapporo~060-0817,~Japan}}

\date{}

\begin{document}

\begin{flushright}
  KOBE-TH-23-02 \\ KU-PH-034 \\ EPHOU-23-007 
\end{flushright}

{\let\newpage=\relax \maketitle}

\begin{abstract}
We study the generation number of massless fermions in compactifications recently found in heterotic supergravity. The internal spaces are products of two-dimensional spaces of constant curvature, and the standard embedding is not assumed. 
The generation number is constrained by the equations of motion and the Bianchi identity. 
In the case that the Euler characteristics of the three internal submanifolds are $(2,-2,-2)$, the generation number is three or less. We also show that three-generation solutions exist for arbitrary Euler characteristics of the negatively curved 2-manifolds, and we present some explicit solutions.
\end{abstract}

\section{Introduction}

The origin of three-generation quarks and leptons is one of the mysteries of the standard model.
There may be an answer in compactifications of the extra dimensions in the superstring theories.
Since the mid-1980s, supersymmetric compactifications like Calabi-Yau manifolds~\cite{Candelas:1985en} have mainly been considered in this context.
In particular, toroidal orbifold compactifications, which are singular limits of certain Calabi-Yau compactifications, have been well studied to derive three-generation models; see, e.g., Refs.~\cite{Ibanez:1987sn,Libanov:2000uf,Frere:2000dc,Abe:2008fi,Abe:2008sx,Abe:2013bca,Abe:2015yva,Sakamoto:2020pev,Kobayashi:2022xsk,Kobayashi:2022tti,Imai:2022bke,Hoshiya:2020hki,Kikuchi:2022lfv,Kikuchi:2022psj}.

However, supersymmetry has not been discovered by collider experiments, so low-energy supersymmetry may not be a necessary condition.
If supersymmetry is not assumed, we have to solve the equations of motion to find compact spaces (for discussions on the relation between the supersymmetry conditions and the equations of motion, see Refs. \cite{Ivanov:2009rh,Lechtenfeld:2010dr}).
This is difficult in general since they are second-order nonlinear partial differential equations.

Recently, nonsupersymmetric exact solutions of equations of motion in heterotic supergravity were found \cite{Tsuyuki:2021xqu}.
They are direct products of Minkowski spacetime and 2-manifolds: $S^2\times T^2\times H^2/\Gamma$, $S^2\times S^2\times H^2/\Gamma$, and $S^2\times H^2/\Gamma\times H^2/\Gamma$, where $S^2$, $T^2$, and $H^2/\Gamma$ denote spaces of positive, zero, and negative constant curvature (a sphere, a torus, and a compact hyperbolic manifold), respectively \cite{stillwell1995geometry,Arefeva:1985yn,Orlando:2010kx}. 
These are Riemann surfaces with genus 0, 1, and $\geq 2$, respectively.
There are many advantages of considering such manifolds: the equations of motion are simplified to algebraic equations, the curvatures of compact dimensions are discretely fixed, the scale of compact space volumes is determined by $\alpha'$ (except $T^2$), the standard embedding does not need to be assumed, the 3-form $H$ field is naturally zero, and the Yukawa couplings may be calculated by applying previous studies on two-dimensional compactifications ($S^2$ was studied in Refs.~\cite{Conlon:2008qi,Imai:2019xdy} and $H^2/\Gamma$ in Ref.~\cite{Honda:2020ckp}). 
Product spaces of 2-manifolds can satisfy the Bianchi identity for $H$ without the standard embedding \cite{Witten:1984dg}.
In this case, the generation number of fermions $N_{\text{gen}}$ is determined by gauge fluxes \cite{Bars:1985nz} and not by the Euler number $\chi$ only; the well-known relation $N_{\text{gen}}=|\chi/2|$ does not hold. 
In this paper,
we study the possible $N_{\text{gen}}$ in these compactifications by considering both the equations of motion and the Bianchi identity.
In Sec.~\ref{sec:com}, we summarize these equations under the assumption that the internal space is a product of two-dimensional spaces of constant curvature. 
In Sec.~\ref{sec:ngen}, we give the explicit relation between $N_{\text{gen}}$ and fluxes. By the relation and an equation of motion, we obtain $N_{\text{gen}}=0$ for $S^2\times T^2\times H^2/\Gamma$.
In Sec.~\ref{sec:ran}, we consider the simplest case that the Euler characteristics of $H^2/\Gamma$ are $-2$. In this case, we show that $N_{\text{gen}}\leq 3$ for the $S^2\times H^2/\Gamma\times H^2/\Gamma$. On the other hand, a larger $N_{\text{gen}}$ (in our scope, 9) is possible for $S^2\times S^2\times H^2/\Gamma$.
Finally, in Sec.~\ref{sec:thr}, we prove that the $N_{\text{gen}}=3$ solutions always exist for both $S^2\times S^2\times H^2/\Gamma$ and $S^2\times H^2/\Gamma\times H^2/\Gamma$ with arbitrary Euler characteristics.

\section{Compactification} \label{sec:com}
\subsection{Equations of motion}

We review the compactification found in Ref. \cite{Tsuyuki:2021xqu} and summarize the conditions to be solved in the following sections.
We consider the bosonic part of the Lagrangian for heterotic supergravity \cite{Bergshoeff:1989de,Lechtenfeld:2010dr}:
\begin{align}
L=\sqrt{-g}e^{-2\phi}\left[R+4(\nabla \phi)^2-\frac{1}{12}H_{MNP}H^{MNP}+\frac{\alpha'}{8}R_{MNPQ}R^{MNPQ}-\frac{\alpha'}{8}\textrm{tr}(F_{MN}F^{MN})\right],
\end{align}
where $g$ is the determinant of the metric $g_{MN}$ and the indices run $M,N,P,Q =0,\ldots,9$. $R_{MNPQ}$ is the Riemann tensor, the Ricci scalar $R$ and Ricci tensor $R_{MN}$ are defined as  $R= g^{NQ}R_{NQ} = g^{MP}g^{NQ} R_{MNPQ}$, and $F_{MN}$ is the gauge field strength of the gauge group $E_8\times E_8$.

The equation of motion for the 3-form field $H$ is
\begin{align}
\partial^M(e^{-2\phi}H_{MNP})=0. \label{ehmn}
\end{align}
We take the dilaton $\phi$ as a constant and assume that $H$ vanishes,
\begin{align}
\partial_M \phi=0,\quad H_{MNP}=0, \label{edp}
\end{align}
and then the equation of motion for $H$ [Eq.~(\ref{ehmn})] is satisfied. Such a configuration with unbroken supersymmetry leads to Ricci-flat compactification \cite{Candelas:1985en}. In this paper, we do not assume low-energy supersymmetry. The other equations of motion \cite{Becker:2009df,Lechtenfeld:2010dr} become
\begin{align}
R+\frac{\alpha'}{8}R_{MNPQ}R^{MNPQ}-\frac{\alpha'}{8}\textrm{tr}(F_{MN}F^{MN}) &=0,\label{eral}\\
R_{MN}+\frac{\alpha'}{4}R_{MPQR}{R_N}^{PQR} -\frac{\alpha'}{4}\textrm{tr}(F_{MP}{F_N}^{P}) &=0, \label{ermn}\\
\nabla_M F^{MN}+[A_{M},F^{MN}] &=0. \label{efm}
\end{align}
By multiplying Eq. (\ref{ermn}) by $g^{MN}$ and comparing with Eq. (\ref{eral}), the dilaton equation becomes much simpler:
\begin{align}
R=0. \label{er0}
\end{align}

In addition to the above equations of motion, the curvature 2-form $\mathcal{R}$
and the gauge field 2-form $F$ 
have to satisfy the Bianchi identity
\begin{align}
0=dH=\frac{\alpha'}{4}(\textrm{tr} \mathcal{R}\wedge \mathcal{R} - \textrm{tr} F\wedge F) \label{edh}
\end{align}
for the Green-Schwarz anomaly cancellation mechanism \cite{Green:1984sg}.

\subsection{Curvatures and gauge field configurations}
We are going to solve Eqs. (\ref{ermn})--(\ref{edh}). We assume that the ten-dimensional spacetime $M^{10}$ is a product of four manifolds:
\begin{align}
M^{10}=M_0\times M_1\times M_2\times M_3, \label{em10}
\end{align}
where $M_0$ is the four-dimensional Minkowski spacetime and $M_i\ (i=1,2,3)$ are two-dimensional spaces of constant curvature. Note that the antisymmetric tensor field $H_{MNP}$ is naturally zero for two-dimensional $M_i$. The metric of $M^{10}$ is block diagonal and depends only on the coordinates of corresponding submanifolds: $g_{mn}^{(i)}=g_{mn}^{(i)}(x^{(i)})$, where the indices $m$ and $n$ are tangent to $M_i$. The nonzero Riemann tensor components are then
\begin{align}
R_{mnpq}^{(i)} &=\lambda_i (g_{mp}^{(i)}g_{nq}^{(i)}-g_{mq}^{(i)}g_{np}^{(i)}), \label{erm}
\end{align}
where $\lambda_i$ is a constant sectional curvature. The manifold with $\lambda_i>0$ is a sphere $S^2$, $\lambda_i=0$ is a torus $T^2$, and $\lambda_i<0$ is a compact hyperbolic manifold $H^2/\Gamma$ \cite{stillwell1995geometry,Arefeva:1985yn,Orlando:2010kx}.

For the gauge field strength $F_{MN}=F_{AMN}T_A$ ($T_A$ are generators of $E_8\times E_8$), we assume that it is also block diagonal for $M$ and $N$ and nonzero only for $U(1)_A$ components $(A=1,2,\ldots,11)$. The range of $A$ is set to leave the gauge group $SO(10)$ in $E^8\times E^8\supset SO(10)\times U(1)_{1}\times\dots\times U(1)_{11}$. 
For each block of $F_{AMN}$, we assume the Freund-Rubin configuration \cite{Freund:1980xh}
\begin{align}
F_{Ai,mn} &=\sqrt{g_i}f_{Ai} \epsilon_{mn}^{(i)}, \label{efmn}
\end{align}
where $g_i=\textrm{det}(g_{mn}^{(i)})$, $f_{Ai}$ is a constant (sometimes called a flux density \cite{Brown:2013mwa}), and $\epsilon_{mn}^{(i)}$ is a Levi-Civita symbol for $M_i$ (for example, $\epsilon_{45}^{(1)}=-\epsilon_{54}^{(1)}=1$).

The configuration of Eq. (\ref{efmn}) satisfies the equation of motion for $F_{MN}$ [Eq.~(\ref{efm})].
The other equations of motion [Eqs.~(\ref{ermn}) and (\ref{er0})] and the Bianchi identity (\ref{edh}) are much simplified (we take $\alpha'=2$ for simplicity) \cite{Tsuyuki:2021xqu}: 
\begin{align}
\lambda_1+\lambda_2+\lambda_3 &= 0,   \label{esum}\\
\lambda_i +\lambda_i^2 -\sum_{A=1}^{11}f_{Ai}^2&= 0,   \label{elamb} \\
\sum_{A=1}^{11} f_{Ai}f_{Aj} &= 0 \quad (i\neq j). \label{ebi}
\end{align}
The first equation of motion (\ref{esum}) does not allow curvatures $\lambda_i$ to be all positive or all negative.
If $\lambda_i=0$, the second equation of motion (\ref{elamb}) necessitates $f_{Ai}=0$ for all $A$. In this case, the generation number is zero as we will see in the following section. We consider the cases with $\lambda_i\neq 0$.
We set $\lambda_1\geq \lambda_2\geq \lambda_3$ without loss of generality, and then $M_1=S^2$ and $M_3=H^2/\Gamma$. 
For the second submanifold $M_2$, both $S^2$ and $H^2/\Gamma$ are possible.

\subsection{Euler characteristics and flux quantization}

Equations (\ref{esum}), (\ref{elamb}), and (\ref{ebi}) are actually integer equations, as we will see below.
The sphere $S^2$ and the compact hyperbolic manifold $H^2/\Gamma$ are compact, boundaryless surfaces. Their genera $g$ are 0 and $g\geq 2$, respectively.\footnote{Any Riemann surface of genus $g\geq 2$ can be expressed as a quotient $H^2/\Gamma$. For example, the surface of $g=2$ is obtained by a discrete group $\Gamma$ that has an octagonal fundamental region in the hyperbolic plane $H^2$. See Refs. \cite{stillwell1995geometry,Arefeva:1985yn,Orlando:2010kx} for more details.} The Euler characteristics $\chi$ and genera are related by $\chi = 2-2g$. By the Gauss-Bonnet formula for compact and boundaryless surfaces, we obtain
\begin{align} 
\int_{M_i}\frac{R^{(i)}}{2}\sqrt{g_i}d^2x=\textrm{vol}(M_i)\lambda_i=2\pi\chi_i, \label{erv}
\end{align}
where $R^{(i)}$ is the Ricci scalar of $M_i$ and $\chi_i$ is its Euler characteristic. 
In addition, the gauge field strength satisfies the flux quantization condition \cite{Witten:1984dg}
\begin{align} 
\int_{M_i} F_{Ai} = \textrm{vol}(M_i) f_{Ai} = 2\pi n_{Ai}, \label{efv}
\end{align}
where $n_{Ai}$ is an integer.
For $\chi_i\neq 0$, from Eqs. (\ref{erv}) and (\ref{efv}) we find
\begin{align}
f_{Ai}=\frac{n_{Ai}}{\chi_i}\lambda_i. \label{efia}
\end{align}
By substituting Eqs. (\ref{efv}) and (\ref{efia}) into Eqs. (\ref{elamb}) and (\ref{ebi}), the equations of motion and the Bianchi identity become
\begin{align}
&\sum_{A=1}^{11}n_{Ai}^2 = \chi_i^2\left(1+\frac{1}{\lambda_i}\right), \label{elic}\\
&\sum_{A=1}^{11} n_{Ai}n_{Aj} = 0 \quad (i\neq j). \label{eq:bi}
\end{align}
The equations to be solved in this paper are Eqs.~(\ref{esum}), (\ref{elic}), and (\ref{eq:bi}).

\section{Number of generations} \label{sec:ngen}

In this section, we obtain an expression for the generation number $N_{\text{gen}}$ from the fluxes $n_{Ai}$ by applying the general formula in Ref. \cite{Bars:1985nz}. 

We consider the  decomposition of $E_8$ and a gaugino in the 248 representation \cite{Green:2012pqa}:
\begin{align}
E_8 &\supset SO(10)\times SU(4),\\
248 &= (45,1)+(1,15)+(10,6)+(16,4)+(\overline{16},\overline{4})
\end{align}
One generation of the standard model fermions can be in a 16 representation.  
In general, the number of generations depends on two terms proportional to $\text{tr} (F^3)$ and $\text{tr} (F) c_2(\mathcal{R})$, where $c_2$ denotes the second Chern class \cite{Bars:1985nz}.
We consider fluxes in $SU(4)$, and thus $\text{tr} (F)=0$; the second term vanishes. Thus, the number of generations is
\begin{align}
N_{\text{gen}} &= \frac{1}{(2\pi)^3}\frac{1}{6}\left|\int \text{tr}(F^3)\right|.
\end{align} 
To be concrete, we take the basis of the Cartan subalgebra of $\mathfrak{su}(4)$ as
\begin{align}
\begin{split}
T_1&=
\text{diag}[-1,0,0,1],\\
T_2&=
\text{diag}[0,-1,0,1],\\
T_3&=
\text{diag}[0,0,-1,1].    
\end{split}
\end{align}
We consider that $U(1)_A\ (A=1,2,3)$ fluxes are in the $SU(4)$. The other fluxes $U(1)_A\ (A=4,\dots,11)$ in another $E_8$ are not relevant for $N_{\text{gen}}$. Then, by using $F=\sum_{A=1}^3 F_AT_A$, the generation number is
\begin{align}
N_{\text{gen}} 
&=\frac{1}{(2\pi)^3}\frac{1}{6}\left|\int 
\text{tr}\{(\text{diag}[-F_1,-F_2,-F_3,F_1+F_2+F_3])^3\}
\right| \notag\\
&=\frac{1}{(2\pi)^3}\frac{1}{6}\left|\int 
\left(\left(\sum_{A=1}^3F_A\right)^3-\sum_{A=1}^3 F_{A}^3\right)\right|. \label{en16}
\end{align}
As an example, we calculate one of the second terms in Eq.~(\ref{en16}) explicitly:
\begin{align}
\frac{1}{(2\pi)^3}\frac{1}{6}\int F_{A}^3
&=\frac{1}{(2\pi)^3}\int _{M_1}F_{A1}\int_{M_2}F_{A2}\int_{M_3}F_{A3} \notag\\
&=\prod_{i=1}^3n_{Ai}.
\end{align}
We have used the flux quantization condition (\ref{efv}).  
The other terms can be obtained similarly. The result is
\begin{align}
N_{\text{gen}}=&\left|\prod_{i=1}^3\sum_{A=1}^3 n_{Ai} -\sum_{A=1}^3 \prod_{i=1}^3n_{Ai}\right|. \label{engen}
\end{align}
We calculate $N_{\text{gen}}$ using this formula in this paper.

If one of the submanifolds $M_i$ is $T^2$, $\lambda_i=0$ and Eq.~(\ref{elamb}) implies $n_{Ai}=0$ for all $A$, then $N_{\text{gen}}=0$.
Thus, nonzero generation numbers are possible only for the compactifications with $S^2\times H^2/\Gamma\times H^2/\Gamma$ or $S^2\times S^2\times H^2/\Gamma$, and we consider these manifolds below.

\section{Ranges of generation numbers} \label{sec:ran}

The generation number $N_{\text{gen}}$ depends on $n_{Ai}$, and $n_{Ai}$ are constrained by the equations of motion (\ref{esum}), (\ref{elic}), and (\ref{eq:bi}). In this section, we discuss the possible ranges of $N_{\text{gen}}$ that satisfy these constraints.

\subsection{$S^2\times H^2/\Gamma\times H^2/\Gamma$ case}

First, we show that the curvature $\lambda_1$ is uniquely determined in this case. 
For the sphere $S^2$, $\lambda_1>0$ and $\chi_1=2$. By substituting it into Eq. (\ref{elic}), $n_{A1}$ have to satisfy
\begin{align}
  \sum_{A=1}^{11}n_{A1}^2 > 4. \label{S2sumn}
\end{align}
To obtain nonzero $N_{\text{gen}}$, at least one of $\{n_{Ai}\}_{A=1,2,3}$ for every $i$ has to be nonzero; $\sum_{i=1}^3 n_{Ai}^2\geq 1$. To satisfy this condition in the hyperbolic regions $\lambda_2<0$, $\lambda_3<0$, Eq. (\ref{elic}) necessitates $\lambda_2<-1$, $\lambda_3<-1$.
By these inequality and Eq. (\ref{esum}), we obtain $\lambda_1=-\lambda_2-\lambda_3>2$. By applying it to Eq. (\ref{elic}), $n_{A1}$ also need to satisfy
\begin{align}
\sum_{A=1}^{11}n_{A1}^2 < 6.
\end{align}
Thus, there is only one possibility,
\begin{align}
\sum_{A=1}^{11}n_{A1}^2 =5,
\end{align}
and it corresponds to $\lambda_1=4$.

On the other hand, the conditions $\lambda_2<0$, $\lambda_3<0$ mean that
\begin{align}
\sum_{A=1}^{11}n_{Ai}^2<\chi_i^2 \quad (i=2,3). \label{eq:naiup}
\end{align}
In general, $|\chi_i|\ (i=2,3)$ and the fluxes can be arbitrarily large. The generation number can be (almost) any integer in our compactifications.
However, it would be natural that the Euler characteristics and fluxes are not so large.

To be concrete, we consider the simplest case $\chi_2=\chi_3=-2$ in this section. We can show that $N_{\text{gen}}\leq 3$ in this case.
From Eq. (\ref{eq:naiup}), there are three possibilities: $\sum_A n_{Ai}^2=1,2,3$ ($i=2,3$). These correspond to $\lambda_2,\lambda_3 = -4/3,-2,-4$. Only one pattern of curvatures can satisfy the equation of motion (\ref{esum}):
\begin{align}
\lambda_1=4, \quad \lambda_2=\lambda_3=-2.
\end{align}
Equivalently, the fluxes need to satisfy
\begin{align}
\sum_{A=1}^{11}n_{A1}^2=5, \ \sum_{A=1}^{11}n_{A2}^2=\sum_{A=1}^{11}n_{A3}^2=2. 
\label{enval}
\end{align}
Then, $n_{A1}=(2,1,\vec{0})$ or $(1,1,1,1, 1,\vec{0})$, $n_{A2}=(1,1,\vec{0})$, $n_{A3}=(1, 1,\vec{0})$ or their permutations and the
signs of each component can be changed.
The fluxes that contribute to $N_{\text{gen}}$ are $A=1,2,3$, and these fluxes satisfy 
\begin{align}
\left|\sum_{A=1}^3 n_{A1}\right|\leq 3,\ \left|\sum_{A=1}^3 n_{Ai}\right|\leq 2 \ (i=2,3). \label{eq:naup}
\end{align}

The Bianchi identity (\ref{eq:bi}) restricts $N_{\text{gen}}$ further. It requires each column of $n_{Ai}$ to be orthogonal to the others. Since only two components of columns $i=2,3$ are nonzero, $n_{A1}n_{A2}n_{A3}=0$ for every $A$. 
Thus, the second term of $N_{\text{gen}}$ in Eq. (\ref{engen}) vanishes. 

The latter inequality of Eq.~(\ref{eq:naup}) is satisfied in three cases: $\left|\sum_{A=1}^3 n_{Ai}\right|= 0 \text{ or } 2$ for one of $i\in\ \{2,3\}$, or $\left|\sum_{A=1}^3 n_{A2}\right|= \left|\sum_{A=1}^3 n_{A3}\right|=1$.
In the first case, $N_{\text{gen}}=0$.  
In the second case, $\left|\sum_{A=1}^3 n_{Ai}\right|$ cannot be 2 for both $i=2,3$ by the Bianchi identity. For example, if $(n_{A3})=(1,1,0,\dots)$, the Bianchi identity necessitates $(n_{A2})=(0,0,1,\dots)$ and $(n_{A1})=(0,0,1,\dots)$ or $(\pm1,\mp1,1,\dots)$, and then $N_{\text{gen}}\leq 2$.
In the last case,
\begin{align}
N_{\text{gen}}=\left|\sum_{A=1}^3 n_{A1}\right|\cdot 1\cdot1\leq 3.
\end{align}
The equality $N_{\text{gen}}=3$ can hold and an example is shown in Solution~1 of Table~\ref{tab:sol}. Thus, the maximal generation number is three when $\chi_2=\chi_3=-2$.

\subsection{$S^2\times S^2\times H^2/\Gamma$ case}
Next, in the $M_2=S_2$ case the sectional curvatures are not uniquely determined, and larger $N_{\text{gen}}$ are possible. The curvatures satisfying the equations of motion (\ref{esum}) and (\ref{elic}) are
\begin{align}
(\lambda_1,\lambda_2,\lambda_3)=(2,2,-4),(1,1,-2),
\left(\frac{4}{3},\frac{2}{3},-2\right),
\left(1,\frac{1}{3},-\frac{4}{3}\right),
\left(\frac{2}{3},\frac{2}{3},-\frac{4}{3}\right).
\end{align}
For example, let us consider the case $(\lambda_1,\lambda_2,\lambda_3)=(\frac{2}{3},\frac{2}{3},-\frac{4}{3})$ and $\chi_3=-2$.
In this case, the fluxes are
\begin{align}
\sum_{A=1}^{11}n_{A1}^2=\sum_{A=1}^{11}n_{A2}^2=10, \ \sum_{A=1}^{11}n_{A3}^2=1. 
\label{enval2}
\end{align}
These equations constrain the fluxes $n_{A1}$ and $n_{A2}$ to be permutations of $(3,1,\vec{0})$, $(2,2,1,1,\vec{0})$, $(2,1,1,1,1,1,1,\vec{0})$ or $(1,1,1,1,1,1,1,1,1,1,0)$ and $n_{A3}$ to be a permutation of $(1,\vec{0})$ (the signs of each component can be changed). We have checked all the cases that also satisfy the Bianchi identity (\ref{eq:bi}). We found that $N_{\text{gen}}\leq 9$ and $N_{\text{gen}}\neq 5,7$. The fluxes that give $N_{\text{gen}}=9$ are, for example, 
\begin{align}
(n_{Ai})=
\begin{pmatrix}
    3&0&0\\
    0&3&0\\
    0&0&1\\
    1&0&0\\
    0&1&0\\
    \vec{0}&\vec{0}&\vec{0}
\end{pmatrix},
\end{align}
where $\vec{0}$ is a six-component zero vector. The fluxes that give $N_{\text{gen}}=3$ are shown in Solution 4 of Table~\ref{tab:sol}.

\tabcolsep = 2pt
\begin{table}[tb]
\centering
\begin{tabular}{ccccccc}
\hline \hline
& Solution 1& Solution 2& Solution 3& Solution 4&Solution 5& Solution 6\\ \hline
$\chi_2$ &$-2$&$-2$&$-4$  & 2& 2& 2\\
$\chi_3$ &$-2$& $-4$&$-4$&$-2$&$-4$&$-6$\\
$(n_{Ai})$
&$\begin{pmatrix}2&0&0\\ 1&0&0\\ 0&1&1\\ 0&1&-1\\ 0&0&0\\ 0&0&0\\ 0&0&0\\ 0&0&0\\ 0&0&0\\ 0&0&0\\ 0&0&0\\ \end{pmatrix}$
& $\begin{pmatrix}2&0&-1\\ 1&0&2\\ 0&1&0\\ 0&1&0\\ 0&0&0\\ 0&0&0\\ 0&0&0\\ 0&0&1\\ 0&0&1\\ 0&0&1\\ 0&0&0\end{pmatrix}$
&$\begin{pmatrix}2&0&-1\\ 1&0&2\\ 0&1&0\\ 0&2&0\\ 0&1&0\\ 0&1&0\\ 0&1&0\\ 0&0&1\\ 0&0&1\\ 0&0&1\\ 0&0&0\end{pmatrix}$
&$\begin{pmatrix}3&0&0\\ 0&1&0\\ 0&0&1\\ 1&0&0\\ 0&3&0\\ 0&0&0\\ 0&0&0\\ 0&0&0\\ 0&0&0\\ 0&0&0\\ 0&0&0\\ \end{pmatrix}$
& $\begin{pmatrix}3&0&0\\ 0&1&0\\ 0&0&1\\ 1&0&0\\ 0&3&0\\ 0&0&1\\ 0&0&1\\ 0&0&1\\ 0&0&0\\ 0&0&0\\ 0&0&0\\ \end{pmatrix}$
& $\begin{pmatrix}3&0&0\\ 0&1&0\\ 0&0&1\\ 1&0&0\\ 0&3&0\\ 0&0&2\\ 0&0&2\\ 0&0&0\\ 0&0&0\\ 0&0&0\\ 0&0&0\\ \end{pmatrix}$

 \\
\hline \hline
\end{tabular}
\caption{Samples of three-generation solutions. These Euler characteristics $\chi_i$ and fluxes $(n_{Ai})$ satisfy the equations of motion (\ref{esum}) and (\ref{elic}), the Bianchi identity (\ref{eq:bi}), and $N_{\text{gen}}=3$ [$N_{\text{gen}} $ is given in Eq. (\ref{engen})]. Solutions 1--3 are for $S^2\times H^2/\Gamma\times H^2/\Gamma$ cases, and Solutions 4--6 are for $S^2\times S^2\times H^2/\Gamma$ cases.} 
\label{tab:sol}
\end{table}

\section{Three-generation solutions} \label{sec:thr}

In this section, we show that three-generation solutions exist for arbitrary Euler characteristics of $H_2/\Gamma$, in both the $M_2=S_2$ and $H_2/\Gamma$ cases. Samples of the solutions are summarized in Table~\ref{tab:sol}.

\subsection{$S^2\times H^2/\Gamma\times H^2/\Gamma$ case}
In this case, we search for the solution with $\lambda_1=4$, $\lambda_2=\lambda_3=-2$.
In a special case with $\chi_2=\chi_3=-2$, we can find a solution as shown in Table~\ref{tab:sol} (Solution 1). In the other cases, we can set $\chi_3\leq -4$ without loss of generality, and we can choose
\begin{align}
(n_{Ai})=
\begin{pmatrix}
    2&0&-1\\
    1&0&2\\
    0&1&0\\
    \vec{0}&\vec{n}_2'&\vec{0}\\
    \vec{0}&\vec{0}&\vec{n}_3'
\end{pmatrix},
\end{align}
where $\vec{n}_2'=(n_{42},\dots,n_{72})$ and $\vec{n}_3'=(n_{83},\dots,n_{11,3})$ are four-component vectors. Such fluxes satisfy the Bianchi identity (\ref{eq:bi}) and the three-generation condition. 
From Eq.~(\ref{elic}), the curvatures $\lambda_2=\lambda_3=-2$
can be obtained from the fluxes satisfying
\begin{align}
\sum_{A=4}^7 n_{A2}^2 &=\frac{\chi_2^2}{2}-1,\\
\sum_{A=8}^{11} n_{A3}^2 &=\frac{\chi_3^2}{2}-5. 
\end{align}
Since $\chi_2\leq -2$, $\chi_3\leq -4$ and both are even numbers, the right-hand sides are positive integers. According to Lagrange's four-square theorem (see Ref. \cite{Eastwood} for a recent discussion), every positive integer can be written as the sum of at most four squares. Interestingly, we have four integer parameters on the left-hand sides, so the solutions of $\lambda_2=\lambda_3=-2$ always exist. 

For example, if $\chi_3=-4$, then $\vec{n}_3'=(1,1,1,0)$.
In Table~\ref{tab:sol}, we show three explicit solutions with $(\chi_2,\chi_3)=(-2,-2),\ (-2,-4),\ (-4,-4)$.

\subsection{$S^2\times S^2\times H^2/\Gamma$ case}
To satisfy the equation of motion (\ref{esum}), we search for the solution with $\lambda_1=\lambda_2=2/3$, $\lambda_3=-4/3$. The curvatures $\lambda_1=\lambda_2=2/3$ can be obtained from the fluxes satisfying
\begin{align}
\sum_{A=1}^{11}n_{A1}^2=\sum_{A=1}^{11}n_{A2}^2=10.
\end{align}
Then, we choose
\begin{align}
(n_{Ai})=
\begin{pmatrix}
    3&0&0\\
    0&1&0\\
    0&0&1\\
    1&0&0\\
    0&3&0\\
    \vec{0}&\vec{0}&\vec{n}_3''
\end{pmatrix},
\end{align}
where $\vec{n}_3''=(n_{63},\dots,n_{11,3})$ is a six-component vector. The Bianchi identity (\ref{eq:bi}) and the three-generation condition are satisfied. From Eq.~(\ref{elic}), the condition $\lambda_3=-\frac{4}{3}$ becomes 
\begin{align}
\sum_{A=6}^{11} n_{A3}^2 &=\frac{\chi_3^2}{4}-1.
\end{align}
If $\chi_3=-2$, the right-hand side is 0, and thus $\vec{n}_3''=\vec{0}$. If $\chi_3\leq -4$, the right-hand side is a positive integer, so there is a $\vec{n}_3''$ that satisfies this equation by Lagrange's four-square theorem. Thus, the solutions of $N_{\text{gen}}=3$ always exist.

For example, if $\chi_3=-6$, then $\vec{n}_3''=(2,2,0,0,0,0)$. Three sample solutions with $\chi_3=-2,-4,-6$ are shown in Table~\ref{tab:sol}.

\section{Conclusion}
In this paper we have discussed how the generation number $N_{\text{gen}}$ is restricted by both the equations of motion and the Bianchi identity in  $S^2\times H^2/\Gamma\times H^2/\Gamma$ and $S^2\times S^2 \times H^2/\Gamma$. 
In Sec.~\ref{sec:com}, we reviewed the compactification and summarized the equation of motion (\ref{esum}) and (\ref{elic}), and the Bianchi identity (\ref{eq:bi}).
These equations constrain the Euler characteristics $\chi_i$ and the flux quantization numbers $n_{Ai}$.
We saw in Sec.~\ref{sec:ngen} that the generation number $N_{\text{gen}}$ is calculated from the flux quantization numbers $n_{Ai}\,\,(A=1,2,3)$ [Eq.~(\ref{engen})]. 
In Sec.~\ref{sec:ran}, we studied how the generation numbers $N_{\text{gen}}$ satisfying the equations of motion (\ref{esum}) and (\ref{elic}), and the Bianchi identity (\ref{eq:bi}) can be obtained when the Euler characteristics $\chi_i$ of $H^2/\Gamma$ are $-2$. 
We obtained three-generation solutions in both the $S^2\times H^2/\Gamma\times H^2/\Gamma$ and $S^2\times S^2 \times H^2/\Gamma$ cases. In particular, in the case of $S^2\times H^2/\Gamma\times H^2/\Gamma$, the generation number $N_{\text{gen}}$ is at most three. In Sec.~\ref{sec:thr} we showed that three-generation solutions exist for arbitrary Euler characteristics of $H^2/\Gamma$.

These results are interesting for two reasons. One is that, in general, flux quantization numbers $n_{Ai}$ can take arbitrary integers, but they
can be restricted considerably by the equations of motion and the Bianchi identity. The second is that we have obtained three-generation solutions in compactifications without supersymmetry. To derive four-dimensional realistic models, we will study fermion mass hierarchy and flavor mixings as well as $CP$ violation in these compactifications.

\section*{Acknowledgments}
T. T. thanks Kensaku Kinjo and Takayuki Okuda for useful discussions. The authors thank the Yukawa Institute for Theoretical Physics at Kyoto University, where this work was initiated during the YITP-W-22-09 on ``Strings and Fields 2022.'' This work was supported by JSPS KAKENHI Grants No.JP 21J20739 (M. T.) and JP 20J20388 (H. U.).

\bibliographystyle{utphys}
\bibliography{ref}

\end{document}